\documentclass{tMOP2e}

\usepackage{subfigure}
\usepackage{braket}

\theoremstyle{plain}

\theoremstyle{definition}

\theoremstyle{remark}

\begin{document}
\jvol{00} \jnum{00} \jyear{2014} \jmonth{September}
\title{{Bichromatic State-Insensitive Trapping of Cold $^{133}$Cs-$^{87}$Rb Atomic Mixtures} }

\author{M.M. Metbulut$^{\ast}$\thanks{$^\ast$Corresponding author. Email: ucapmmm@ucl.ac.uk
\vspace{6pt}} and F. Renzoni\\\vspace{6pt}  {\em{Department of Physics and Astronomy, University College London, London, UK}}\\
\received{v4.1 released September 2014} }
\maketitle

\begin{abstract}

We investigate simultaneous state-insensitive trapping of a mixture of two different atomic species, Caesium and Rubidium. The magic wavelengths of the Caesium and Rubidium atoms are different, $935.6$ nm and $789.9$ nm respectively, thus single-frequency simultaneous state-insensitive trapping is not possible. We thus identify bichromatic trapping as a viable approach to tune the two magic wavelengths to a common value. Correspondingly, we present several common magic wavelength combinations appropriate for simultaneous state-insensitive trapping of the two atomic species. 


\end{abstract}

\section{Introduction}

Research on multispecies mixtures of cold atoms has been attracting increasing attention over the last decade. Recent progress in the field paved the way to new research areas beyond the experiments with single species. Highlights of these studies include the exploration of phase transitions in optical lattices \cite{phase, lewenstein}, the demonstration of tuning of interatomic interaction between two species \cite{pilch, stan, inouye, ferlaino}, the creation of molecules via photoassociation \cite{pap08, ker, mancini}, sympathetic cooling of one species by another \cite{symp, mudrich, modugno}, the demonstration of quantum degenerate systems \cite{qdc, hadzi, roati}, precision metrology \cite{prm}, quantum computation \cite{ker} and simulation \cite{dm}.

The importance  of optical trapping as a standard tool for the creation and manipulation of ultracold atomic mixtures is undisputed. Optical trapping allows one to produce a potential independent of hyperfine state and spin projection of an atom \cite{odta, odtb}, which is not possible with other trapping mechanisms. Moreover, the use of far-off resonance laser light field for trapping ensures low scattering rates to obtain long coherence times \cite{odta,fort}.
However, optical trapping in general produces a differential AC Stark shift between the different internal states of the atoms.  This presents some drawbacks. 
 For instance, decoherence caused by the differential shift significantly limits the coherence time, which is an important parameter in experiments involving quantum state manipulation \cite{cet}. It also hinders applicability of free space laser cooling techniques. For an optical trap, efficient cooling is not only important for long storage times but also for avoiding motion-induced dephasing of the atoms \cite{cor, yenn}. 

The elimination of the differential AC Stark shift allows one to maintain the internal state of the atom essentially unaffected by the trapping light. State-insensitive trapping,  first proposed by Katori \emph{et} al. \cite{katori1999optimal}, takes advantage of the combined effect of the couplings induced by the trapping light between the ground state and all the excited states of the multi-level atom. By tuning the trapping light to a specific  wavelength, referred as magic wavelength,  the AC Stark shifts of the upper and lower states of a transition of  interested can be made equal.  The application of state-insensitive trapping was first demonstrated in the context of optical lattice clock for Strontium atoms \cite{katori1999optimal}. It was shown that state-insensitive trapping allows simultaneous use of Doppler cooling and optical trapping. Later, it was demonstrated that optical trapping of Caesium atoms within an optical cavity at magic wavelength provides extended life-time \cite{mckeeverpaper} and makes continuous monitoring of the trapped atoms possible. State-insentitive trapping of Caesium atoms was subsequently demonstrated also in free space \cite{hocaboeing}.

Several investigations have been devoted to state-insensitive trapping of single atomic species.  However, the possibility of state-independent trapping of mixtures of different atomic species has remained unexplored. This is treated in detail in the present work. As a case study, we consider the possibility of state-insensitive trapping of a $^{133}$Cs-$^{87}$Rb mixture.  We notice that monochromatic trapping schemes are not suitable for simultaneous state-independent trapping of these atomic species as the magic wavelengths of Caesium and Rubidium are very different, $935.6$ nm and $789.9$ nm respectively. We thus consider bichromatic trapping that provides some control over  the required magic wavelengths. Bichromatic state-insensitive trapping was proposed for both Rubidium \cite{durham,arora, wang} and Caesium atoms \cite{biz} due to inconvenient magic wavelengths for experimental realization either in terms of frequency range or required power of the light source.
This technique involves two independent lasers with different frequencies; one of the lasers is used as the trap laser and the other one as a control laser to counteract the differential light shift induced by the trap laser. In contrast to monochromatic schemes there are four experimental parameters to control, the intensity and wavelength of both lasers. Therefore, the combined effect of the two lasers may cancel the differential shift between the states of an optical transition of interest for various combinations of parameters, thus leading to state-insensitive trapping over a range of wavelengths. By choosing appropriate relative intensity of the trap and control lasers, state-insensitive trapping can be achieved at a specific value of wavelength combination within that range. Such a tunability is the key element  behind simulataneous state-independent trapping of atomic mixtures. 

In this work we demonstrate theoretically the possibility of bichromatic state-independent trapping of a $^{133}$Cs-$^{87}$Rb mixture, and identify the appropriate magic wavelength combinations for simultaneous state-insensitive trapping of the $6S_{1/2}(F=4)\rightarrow6P_{3/2}(F=5)$ and $5S_{1/2}(F=2)\rightarrow5P_{3/2}(F=3)$ transitions of Caesium and Rubidium atoms respectively. 

\section{Magic wavelength combinations for atomic mixtures}

We first evaluate the AC Stark shifts of the ground and excited states of the transitions $6S_{1/2}(F=4)\rightarrow6P_{3/2}(F=5)$ and $5S_{1/2}(F=2)\rightarrow5P_{3/2}(F=3)$ transitions of Caesium and Rubidium atoms respectively,
as a function of the trap laser wavelength in  presence of the trap and control lasers. Then, by varying the trap and control lasers parameters, we will identify the parameters sets which make the magic wavelength combinations for the two atomic species equal.

We performed calculations including all the hyperfine levels and the corresponding Zeeman sublevels. This is especially important whenever the trapped atoms are prepared in a specific Zeeman sublevel. In presence of laser light the AC Stark shift experienced by the atom for a transition from state $\ket{i}=\ket{F_{i}m_{i}}$ to state $\ket{j}=\ket{F_{j}m_{j}}$ is given by

\begin{equation}
\Delta E_i =(2F_{i}+1)I_{L}\sum_{j}{\frac{3\pi c^{2}A_{ij}}{2\Delta_{0}\omega_{0}^{3}}(2F_{j}+1)(2J_{j}+1)\begin{pmatrix} F_{j} & 1 & F_{i} \\ m_{i} & q & -m_{j} \end{pmatrix}^{2}\begin{Bmatrix} J_{i} & J_{j} & 1 \\ F_{j} & F_{i} & I \end{Bmatrix}^{2}}.
\label{actot}
\end{equation}

where I$_{L}$ is the intensity of the laser, $A_{ij}$ is the Einstein coefficient for the optical dipole transition, $\omega_{0}$ is the trapping laser frequency, $\Delta_{0}$ is the detuning from resonance with the transition of interest and $q$ denotes the polarization of the light field; $q=0$ for linear polarization, and $q=\pm 1$ for $\sigma^{\pm}$ polarization. In our calculations, we considered linear polarization, as used in most experiments, a beam waist of $10$ $\mu$m for both trap and control laser fields and a fixed power of $2$ W for the control laser. Therefore, the intensity ratio of the trap and control lasers is used as variable parameter to tune the magic wavelength combinations. In our calculations we took into account contributions from excited states up to $11$S, $12$P and $12$D for Caesium and $10$S, $10$P and $8$D for Rubidium.

The numerical calculations showed that varying the relative intensity of the two lasers does indeed allow for tuning of the magic wavelength pairs to an extent sufficient to make them equal.  We determined six common magic wavelength combinations for Caesium and Rubidium atomic species. These combinations are listed in Table \ref{table1} together with the required intensity ratios of the lasers and the corresponding trap depths. 
Due to the different polarizability of the states of the Caesium and Rubidium atom, the same experimental parameters produce different trap depths for the two species. The ratio of the trap depths of Caesium and Rubidium atoms increases from $\approx$$2$ to $\approx$$2.8$ as the magic trap wavelength increases from $868.8$ nm to $936$ nm. Therefore,  it was not possible to identify a magic wavelength combination which allows for simultaneous state-insensitive trapping of the two atomic species while maintaining the trap depths for the two species equal.

\begin{table}[h]
\tbl{Magic wavelength combinations and corresponding required intensity ratios for Cs-Rb mixture. The trap depths for Cs and Rb atoms are also reported, for a control laser power of $2$ W and a laser beam waists of $10$ $\mu$m for both lasers.}
{\begin{tabular}{@{}lcccccc}\toprule
$\lambda_{t}$ (nm) & $\lambda_{c}$ (nm) & $I_{t}/I_{c}$
& Rb U (mK)
& Cs U (mK) \\ 
\colrule
 868.8 & 10060 & 0.240 & 2.04 & 4.14  \\
868.94 & 10600 & 0.239 & 2.03 & 4.08  \\
 931 & 2038.8 & 0.939 & $3.89$ & 10.17  \\
932.9 & 1843.5 & 1.408 & 5.31 & 14.18  \\
934 & 1742.4 & 1.999 & 7.08 & 19.25  \\
936 & 1580.9 & 7.832 & 24.36 & 68.95  \\
\botrule
\end{tabular}}

\label{table1}
\end{table}

\begin{figure}[h]
\begin{center}
{
\resizebox*{6cm}{!}{\includegraphics{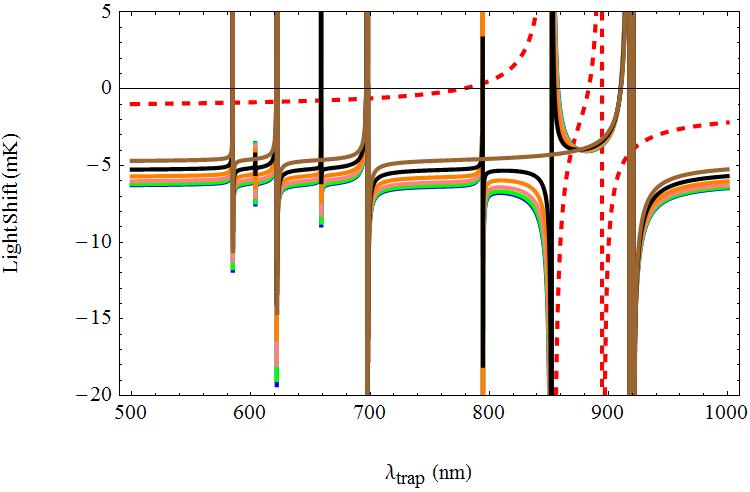}}}

\caption{AC Stark shift of the Caesium D$_2$ line $F_g=4 \to F_e=5$  transition as a function of the trap laser wavelength, with all other parameters corresponding to the magic wavelength combination $\lambda_{t}=868.94$ nm-$\lambda_{c}=10600$ nm, as listed in Table 1. Red dashed line is the ground state $6S_{1/2}F=4$, blue, green, pink, orange, black and brown lines are with $m=0, \pm 1,\pm 2,\pm 3,\pm 4,\pm 5$ Zeeman sublevels, respectively, of the Caesium $6P_{3/2}F=5$ excited state.}
\label{fig1}

\end{center}

\end{figure}

\begin{figure}[h]
\begin{center}
{
\resizebox*{6cm}{!}{\includegraphics{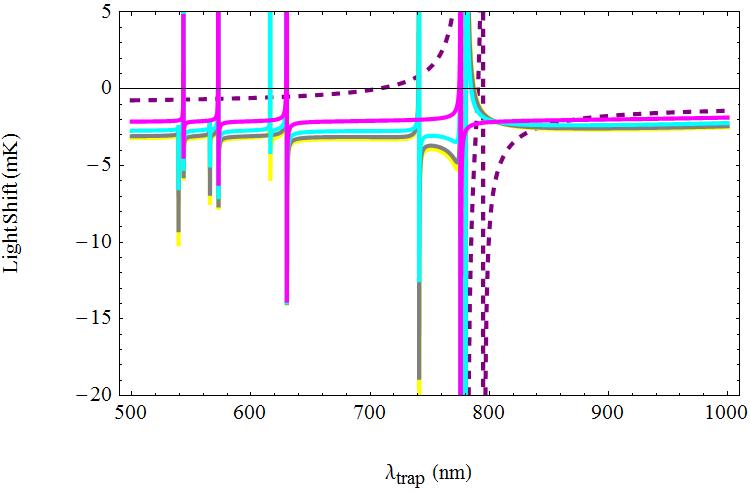}}}

\caption{AC Stark shift of the Rubidium D$_2$ line $F_g=2 \to F_e=3$ transition as function of trap laser wavelength,  with all other parameters corresponding to the magic wavelength combination $\lambda_{t}=868.94$ nm-$\lambda_{c}=10600$ nm, as listed in Table 1. Purple dashed line is the ground state $5S_{1/2}F=2$, yellow, gray, cyan and magenta lines are the $m=0,\pm 1,\pm 2,\pm 3$ Zeeman sublevels, respectively, of the $5P_{3/2}F=3$ Rubidium excited state.}
\label{fig2}

\end{center}

\end{figure}

Out of the six identified magic wavelength combinations, one pair is mostly relevant for practical realizations given current laser technology. This is the magic wavelength pair $\lambda_{t}=868.94$nm--$\lambda_{c}=10600$nm, which corresponds to  diode and CO$_{2}$ laser sources for the trap and control fields, respectively. The light shifts of both atomic species are plotted in figure \ref{fig1} and figure \ref{fig2} as functions of the trap laser wavelength, with the intensity ratio and the control laser wavelength kept fixed to the values corresponding to the magic wavelength combination $\lambda_{t}=868.94$ nm-$\lambda_{c}=10600$ nm, as listed in Table 1. The displayed data demonstrate that at the magic trapping wavelength $\lambda_{t}=868.94$nm the differential light shifts for both atomic species is zero. As shown in both figures, the Zeeman sublevels of the ground state of both atoms remain degenerate, as expected for a linear polarization of the laser fields, whereas they are splitted for the excited state.

\section{Effect of variations in  experimental parameters}

\begin{figure}[h]
\begin{center}
{
\resizebox*{6cm}{!}{\includegraphics{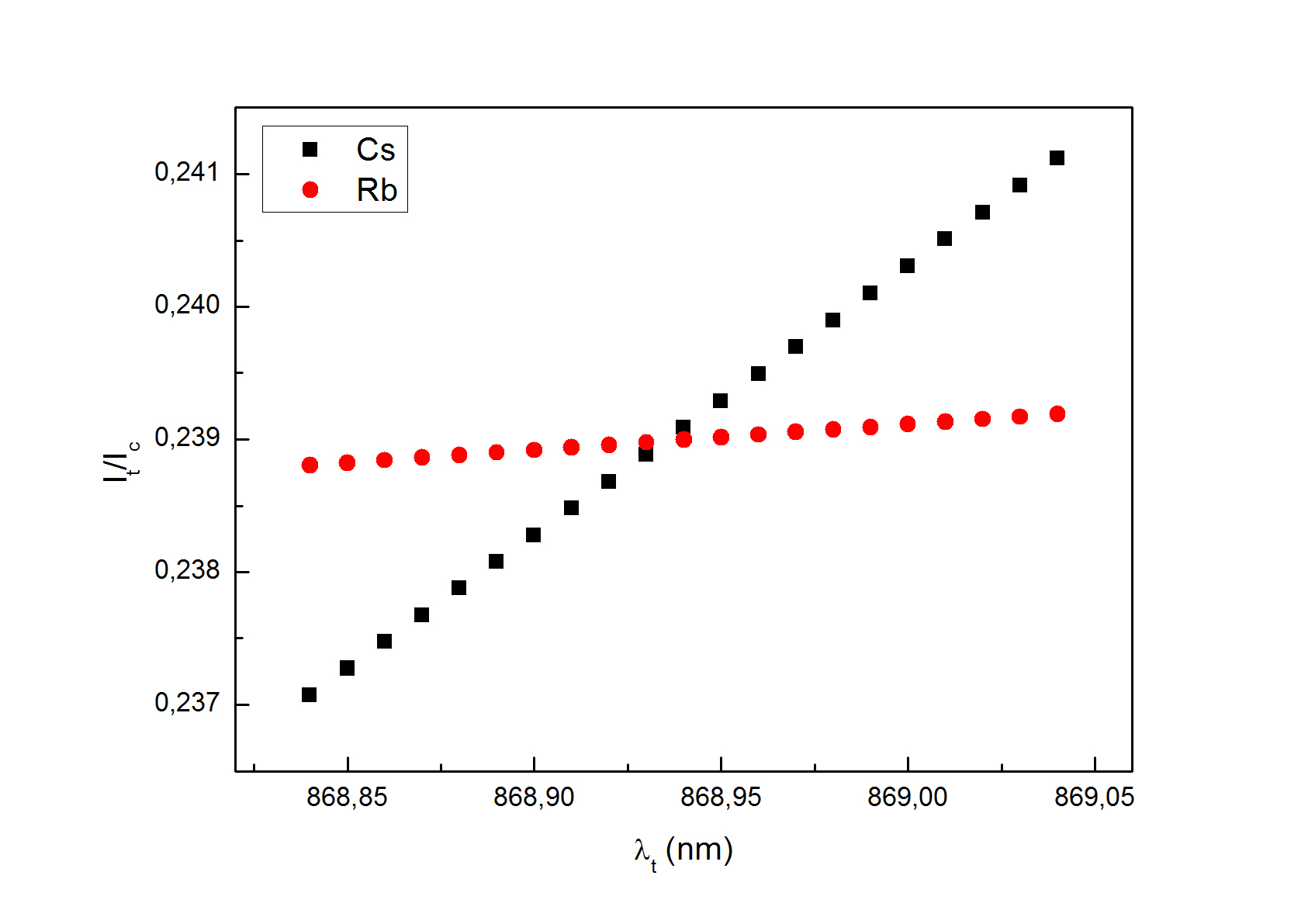}}}
\caption{Trap/control laser intensity ratio required for state-insensitive trapping for Cs and Rb atoms as a function of the trap laser wavelength.}
\label{fig3}

\end{center}
\end{figure}

We also investigated the effect of variations in experimental parameters on  state-insensitive trapping of the  Cs-Rb mixture. First, we examine  how a variation in wavelength of the trapping laser affects the possibility of simultaneous state-independent trapping of the two atomic species.
%
%
We consider variations in trap laser wavelength for the case of the identified magic wavelength combination  $868.94$ nm $-$$10.6$ $\mu$m, which is the most appealing for  experimental implemetations. We calculated the trap/control laser intensity ratio required for state-independent trapping of each individual species, Caesium and Rubidium, as a function of the trap laser wavelength, with results as in figure \ref{fig3}. The data show that a variation in trap laser wavelength cannot be compensated by an appropriate  variation of the trap/control laser intensity ratio. Indeed, away from the trap laser wavelength required for dual-species state-insensitive trapping, the intensity ratio for state-independent trapping of Cs and Rb are different, and  the discrepancy in the required value for the two species increases with increasing variation in the trapping wavelength. In fact, figure \ref{fig3} shows that state-insensitive trapping of Caesium atoms is more sensitive to the variation in the trapping wavelength than the trapping of Rubidium atoms.

The difference between the trap/control intensity ratio required to produce state-independent trapping of Rb and Cs are listed in Table \ref{tab2} for the case of $0.01,0.1$ and $0.5$ nm variation in trap laser wavelength. In agreement with the above analysis for a specific case of magic wavelength combination, the presented data indicates that a small variations in the trap laser wavelength has a substantial effect on the possibility of dual-species state-independent trapping.
The data given in table \ref{tab2} provide a more quantitative indication on the stability requirements for the laser wavelength to maintain state-independent trapping for the mixture. For example, a $0.01$ nm variation in the trapping laser wavelength corresponds to negligible change in the required intensity ratio of the lasers for the two magic wavelength combinations with $\lambda_c=10.6\mu$m. This guarantees that dual-species state-independent trapping will be preserved for wavelength variations below $\pm$$0.01$ nm.

\begin{table}[h]
\tbl{Difference between the required trap/control lasers intensity ratio of the Caesium and Rubidium atoms for maintaining state-insensitive trapping in case of $\Delta\lambda_{t}=0.01,0.1$ and $0.5$ nm variation in trap laser wavelength.}
{\begin{tabular}{@{}lcccccc}\toprule
$\lambda_{t}$ (nm)-$\lambda_{c}$ (nm) & $\Delta$$I_{t}/I_{c}$
for $\Delta\lambda_{t}=0.01$ nm
& $\Delta$$I_{t}/I_{c}$
for $\Delta\lambda_{t}=0.1$ nm & $\Delta$$I_{t}/I_{c}$
for $\Delta\lambda_{t}=0.5$ nm \\ 
\colrule
 868.8-10060 & $3.875\times10^{-4}$ & $6.303\times10^{-4}$ & $8.430\times10^{-3}$  \\
868.94-10600 & $1.855\times10^{-4}$ & $1.734\times10^{-3}$ & $8.870\times10^{-3}$  \\
 931-2038.8 & $2.540\times10^{-3}$ & $5.013\times10^{-2}$ & 0.234 \\
932.9-1843.5 & $7.785\times10^{-3}$ & $5.029\times10^{-2}$ & 0.217 \\
934-1742.4 & $9.355\times10^{-3}$ & $8.628\times10^{-2}$ & 0.376  \\
936-1580.9 & $1.062\times10^{-1}$ & $9.847\times10^{-1}$ & 3.334  \\
\botrule
\end{tabular}}
\label{tab2}
\end{table}

\begin{figure}[h]
\begin{center}
{
\resizebox*{6cm}{!}{\includegraphics{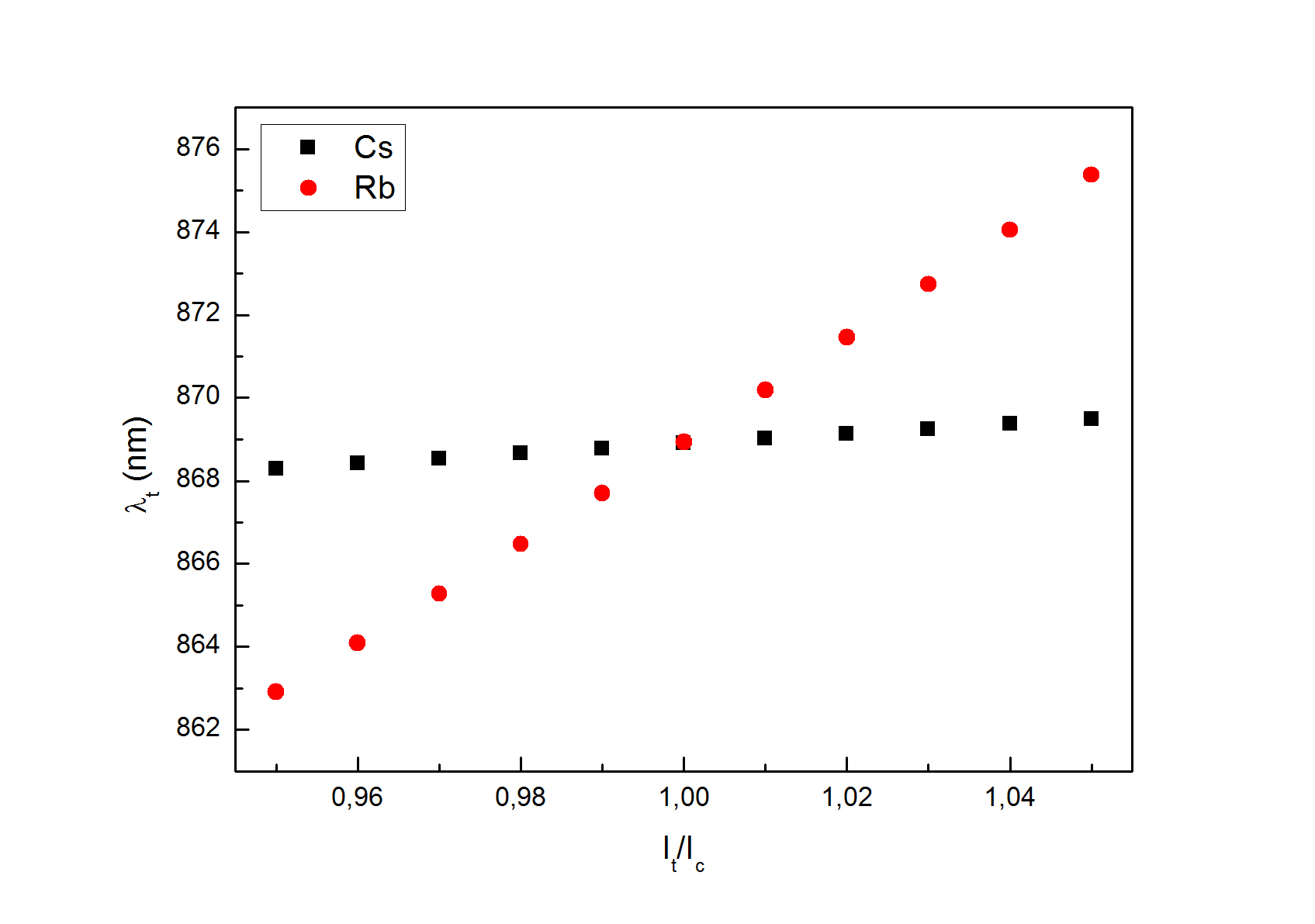}}}
\caption{Trap laser wavelength for state-independent trapping of Caesium and Rubidium atoms as a function of the trap/control laser intensity ratio. All other parameters correspond to the magic wavelength combination $868.94-10600$ nm.}
\label{fig4}

\end{center}
\end{figure}

Next, we analyze the effect of variations in another important parameter, the laser intensity. By considering intensity variations of the trap and control lasers, we investigated the deviations of the corresponding magic trap laser wavelengths of the two atomic species. 
The effect of intensity variations in trap/control laser intensity ratio is illustrated in figure \ref{fig4} for the case of magic wavelength combination of $868.94-10600$ nm. The data indicates that a variation in the intensity ratio leads to different magic wavelength for the two atomic species, with the magic trapping wavelength of the Rb species showing greater deviation than that of the Cs atoms. 

We evaluated the effect of variations of the lasers intensity ratio for all identified magic wavelength combinations. For $0.1$ and $0.5\%$ variation in the trap/control laser intensity ratio, the corresponding magic trap laser wavelengths of the Caesium and Rubidium are listed in table \ref{tab3}. Larger intensity variations cause a larger deviation of the Rb and Cs magic wavelengths from their common magic wavelength.

\begin{table}[h]
\tbl{List of common magic wavelength combinations and  the deviated magic trap laser wavelengths of Cs and Rb for $0.1$ and $0.5\%$ variation in intensity ratios from the value leading to dual-species state-independent trapping. The date refer to a  control laser power of $2$ W and laser beam waists of $10$ $\mu$m for both lasers.}
{\begin{tabular}{@{}lcccccc}\toprule
$\lambda_{t}$ (nm)-$\lambda_{c}$(nm) & $\lambda_{t}$(nm) of Cs-Rb for $\Delta$$I_{t}/I_{c}=0.1\%$
& $\lambda_{t}$(nm) of Cs-Rb for $\Delta$$I_{t}/I_{c}=0.5\%$  \\ 
\colrule
 868.8-10060 & 868.73-868.67 & 868.68-868.18   \\
868.94-10600 & 868.89-868.80 & 868.84-868.30   \\
 931-2038.8 & 931.02-930.64 & 931.01-929.20 &   \\
932.9-1843.5 & 932.94-932.52 & 932.92-931.05   \\
934-1742.4 & 934.03-933.62 & 934.02-932.10  \\
936-1580.9 & 936.04-935.06 & 936.03-934.03  \\
\botrule
\end{tabular}}
\label{tab3}
\end{table}

We notice that the differential deviation between the magic wavelengths of the two atoms is less than $1$ nm for $0.1\%$ variation in the trap/control lasers intensity ratio for all identified magic wavelength combinations as shown in table \ref{tab3}. Indeed, $0.1\%$ variation in the intensity of the lasers for magic wavelength pair of $868.94$ nm $-$$10.6$ $\mu$m leads to $0.1$ nm shift between the corresponding magic wavelengths of the two atoms. Hence, for a typical linewidth of 1 nm for the trap lasers, we can assume that intensity variations below $0.1\%$ will not affect the simultaneous state-insensitive trapping of the two atomic species. 


\section{Conclusions}

In this work we explored the possibility of simultaneous state-independent trapping of two  different atomic species. We considered specifically a Cs-Rb mixture as case study. The magic wavelengths of the Caesium and Rubidium atoms are different, $935.6$ nm and $789.9$ nm respectively, thus single-frequency simultaneous state-insensitive trapping is not possible. We identified bichromatic trapping as a viable approach to tune the two magic wavelengths to a common value. Correspondingly, we presented several common magic wavelength combinations appropriate for simultaneous state-insensitive trapping of the two atomic species. In particular, the configuration with 868.94 nm and 10600 nm trap/control laser wavelength is here individuated as the most promising, also thanks to the ease of implementation with currently available laser systems. The effects of unwanted variations in experimental parameters were also discussed.

\section*{Acknowledgements}
We thank Dr Luca Marmugi (UCL) for critical reading the manuscript

\section*{Funding}
This work was supported by EPSRC [grant number EP/H049231/1].


\begin{thebibliography}{20}

\bibitem[1]{phase}
Altman, E.; Hofstetter, W.; Demler, E.; Lukin, M.D. {\em Jour. of Phys.}
{\bf 2003}, {\em 5} (1), 113.

\bibitem[2]{lewenstein}
Lewenstein, M.; Santos, L.; Baranov, M.A.; Fehrmann, H. {\em Phys. Rev. Lett.}
{\bf 2004}, {\em 92} (5) 050401.

\bibitem[3]{pilch}
Pilch, K.; Lange, A.D.; Prantner, A.; Kerner, G.; Ferlaino, F.; N\"agerl, H.C.; Grimm, R. {\em Phys. Rev. A}
{\bf 2009}, {\em 79} (4), 042718.

\bibitem[4]{stan}
Stan, C.A.; Zwierlein, M.W.; Schunck, C.H.; Raupach, S.M.F.; Ketterle, M. {\em Phys. Rev. Lett.}
{\bf 2004}, {\em 93} (14) 143001.

\bibitem[5]{inouye}
Inouye, S.; Goldwin, J.; Olsen, M.L.; Ticknor, C.; Bohn, J. L.; Jin, D.S. {\em Phys. Rev. Lett.}
{\bf 2004}, {\em 93} (18) 183201.


\bibitem[6]{ferlaino}
Ferlaino, F.; Derrico, C.; Roati, G.; Zaccanti, M.; Inguscio, M.; Modugno, G.; Simoni, A. {\em Phys. Rev. A}
{\bf 2006}, {\em 73} (4) 040702.

\bibitem[7]{pap08}
Papp, S.; Pino, J.; Wieman, C. {\em Phys. Rev. Lett.}
{\bf 2008}, {\em 101} (4), 040402.

\bibitem[8]{ker}
Kerman, A.J.; Sage, J.M.; Sainis, S.; Bergeman, T.; DeMille, D. {\em Phys. Rev. Lett.}
{\bf 2004}, {\em 92} (15), 153001.

\bibitem[9]{mancini}
Mancini, M.W.; Telles, G.D.; Caires, A.R.L.; Bagnato, V.S.; Marcassa, L.G. {\em Phys. Rev. Lett.}
{\bf 2004}, {\em 92} (13) 133203.

\bibitem[10]{symp}
Mosk, A.; Kraft, S.; Mudrich, M.; Singer, K; Wohlleben, W.; Grimm, R.; Weidmuller, M. {\em Appl. Phys. B}
{\bf 2001}, {\em 73} (8) 791.

\bibitem[11]{mudrich}
Mudrich, M.; Kraft, S.; Singer, K.; Grimm, R.; Mosk, A.; Weidemuller, M.  {\em Phys. Rev. Lett.}
{\bf 2002}, {\em 88} (25) 253001.

\bibitem[12]{modugno}
Modugno, G.; Ferrari, G.; Roati, G.; Brecha, R.J.; Simoni, A.; Inguscio, M.  {\em Science}
{\bf 2001}, {\em 294} (5545) 1320.

\bibitem[13]{qdc}
Catani, J.; De Sarlo, L.; Barontini, G.; Minardi, F.; Inguscio, M. {\em Phys. Rev. A}
{\bf 2008}, {\em 77} (1) 011603.

\bibitem[14]{hadzi}
Hadzibabic, Z.; Stan, C.A.; Dieckmann, K.; Gupta, S.; Zwierlein, M.W., Gorlitz, A.; Ketterle, W. {\em Phys. Rev. Lett.}
{\bf 2002}, {\em 88} (16) 160401.

\bibitem[15]{roati}
Roati, G.; Riboli, F.; Modugno, G.; Inguscio, M.   {\em Phys. Rev. Lett.}
{\bf 2002}, {\em 89} (15) 150403.

\bibitem[16]{prm}
Zelevinsky, T.; Kotochigova, S.; Ye, J. {\em Phys. Rev. Lett.}
{\bf 2008}, {\em 100} (4), 043201.

\bibitem[17]{dm}
DeMille, D. {\em Phys. Rev. Lett.}
{\bf 2002}, {\em 88} (6), 067901.

\bibitem[18]{odta}
Grimm, R.; Weidemuller, M.; Ovchinnikov, Y.B. {\em Adv. At. Mol. Opt. Phys.}
{\bf 2000}, {\em 42} (95) 95.

\bibitem[19]{odtb}
Takekoshi, T.; Patterson, B.M.; Knize, R.J. {\em Phys. Rev. Lett.}
{\bf 1998}, {\em 81} (23) 5105.

\bibitem[20]{fort}
Miller, J.D.; Cline, R.A.; Heinzen, D.J. {\em Phys. Rev. A}
{\bf 1993}, {\em 47} (6) R4567.

\bibitem[21]{cet}
Cetinbas, M. {\em Jour. of the Phys. A: Math. and Theor.}
{\bf 2009}, {\em 42} (14), 145302.

\bibitem[22]{cor}
Corwin, K.L.; Kuppens, S.J.M.; Cho, D.; Wieman, C.E. {\em Phys. Rev. Lett.}
{\bf 1999}, {\em 83}, 1311.

\bibitem[23]{yenn}
Ye, J.; Kimble, H.J.; Katori, H. {\em Science}
{\bf 2008}, {\em 320} (5884) 1734.


\bibitem[24]{katori1999optimal}
Katori, H.; Ido, T.; Kuwata-Gonokami, M. {\em Jour. of the Phys. Soc. of Japan}
{\bf 1999}, {\em 68} (8), 2479--2482.

\bibitem[25]{mckeeverpaper}
McKeever, J.; Buck, J. R.; Boozer, A. D.; Kuzmich, A.; N\"agerl, H.-C.; Stamper-Kurn, D. M.; Kimble, H. J. {\em Phys. Rev. Lett.}
{\bf 2003}, {\em 90} (13), 133602.

\bibitem[26]{hocaboeing}
Phoontong, P.; Douglas, P.; Wickenbrock, A.; Renzoni, F. {\em Phys. Rev. A}
{\bf 2010}, {\em 82} (4) 013406.

\bibitem[27]{durham}
Griffin P.F., Weatherill K.J.,  MacLeod S.G., Potvliege R.M. and Adams C.S.,  2006 {\em New J. Phys.} {\bf 2006} (8) 11. 

\bibitem[28]{arora}
Arora, B.; Safronova, M.S.; Charles, W.C. {\em Phys. Rev. A}
{\bf 2010}, {\em 82} (2) 022509.

\bibitem[29]{wang}
Wang, J.; Shanlong, G.; Yulong, G.; Yongjie, C.; Baodong, Y.; Jun, H. {\em J. Phys. B: At. Mol. Opt. Phys.}
{\bf 2014}, {\em 47} (9) 095001.

\bibitem[30]{biz}
Metbulut, M.M.; Renzoni, F. {\em J. Mod. Opt.}
{\bf 2015}, {\em 00} (0) 00000.



\end{thebibliography}
\end{document}